\documentclass[sigconf,natbib=true,anonymous=false]{acmart}
\usepackage{multirow}
\usepackage{graphics}
\usepackage{arydshln}

\usepackage{colortbl}
\usepackage{graphicx}
\usepackage[table]{xcolor}
\usepackage{diagbox}
\usepackage{tabularx}              
\usepackage{booktabs}              
\usepackage{array}                 
\usepackage{kotex}[cjk]
\usepackage{subcaption}
\usepackage{tabularx} 
\AtBeginDocument{%
  }

\setcopyright{acmlicensed}
\copyrightyear{2018}
\acmYear{2018}
\acmDOI{XXXXXXX.XXXXXXX}
\acmConference[Conference acronym 'XX]{Make sure to enter the correct
  conference title from your rights confirmation email}{June 03--05,
  2018}{Woodstock, NY}
\acmISBN{978-1-4503-XXXX-X/2018/06}

\begin{document}

\title{Rescaling MLM-Head for Neural Sparse Retrieval}

\author{Youngjoon Jang}
\orcid{0009-0005-3591-638X}
\affiliation{%
  \institution{Korea University}
  \city{Seoul}
  \country{South Korea}
}
\email{dew1701@korea.ac.kr}

\author{Seongtae Hong}
\orcid{0009-0002-2073-7731}
\affiliation{%
  \institution{Korea University}
  \city{Seoul}
  \country{South Korea}
}
\email{ghdchlwls123@korea.ac.kr}

\author{Jonah Turner}
\orcid{}
\affiliation{%
  \institution{Independent}
  \city{Seoul}
  \country{South Korea}}
\email{drexalt@gmail.com}

\author{Heuiseok Lim}
\orcid{0000-0002-9269-1157}
\authornotemark[1]
\affiliation{%
  \institution{Korea University}
  \city{Seoul}
  \country{South Korea}}
\email{limhseok@korea.ac.kr}

\renewcommand{\shortauthors}{Jang et al.}

\begin{abstract}
Learned sparse retrieval (LSR) models such as SPLADE have traditionally used BERT-style masked language models as backbone encoders. A natural expectation is that replacing BERT with stronger pretrained encoders should improve retrieval effectiveness. However, we find that under standard SPLADE training recipes, backbones with large MLM-head L2 norms can suffer performance degradation and even training collapse under standard SPLADE training recipes. We identify this failure as a scale mismatch in the MLM head: SPLADE directly uses MLM-head outputs to construct sparse lexical representations, and query-document relevance is computed by an unnormalized dot product over these representations. As a result, an inflated MLM-head scale can amplify sparse activations, distort matching scores, and destabilize contrastive training under common training settings. To address this issue, we introduce a simple initialization-time correction that rescales the MLM-head projection by a constant factor before SPLADE training. This zero-cost adjustment improves training stability without modifying the model architecture or training objective. Across both in-domain and out-of-domain retrieval benchmarks, this simple correction substantially improves large-norm backbones such as ModernBERT and Ettin, turning unstable training runs into competitive sparse retrievers. In several settings, the corrected models further match or surpass the classic BERT-SPLADE baseline. These findings suggest that the bottleneck in adapting pretrained encoders to LSR is not encoder capacity alone, but the calibration of the MLM-head scale used to construct sparse lexical representations.

\end{abstract}



\maketitle

\section{INTRODUCTION}

Learned sparse retrieval (LSR) has emerged as an effective compromise between lexical and neural retrieval. By representing queries and documents as sparse vectors over the vocabulary, LSR models retain the efficiency of inverted-index search while enabling neural term expansion and learned term weighting. Among LSR methods, SPLADE has become a widely adopted architecture, achieving strong retrieval effectiveness while preserving the interpretability and deployability of lexical matching~\cite{bm25, splade, spladev2, spladev3}.

A key property of SPLADE is that it constructs sparse lexical representations through the masked language modeling (MLM) head of a pretrained encoder. Unlike dense retrievers, whose representation space is typically learned through a task-specific projection, SPLADE directly reuses the pretrained vocabulary projection to define the sparse retrieval space. This makes the MLM head central not only to pretraining, but also to retrieval scoring.

This dependence suggests a natural path for improving SPLADE: replacing its conventional BERT backbone with stronger modern encoders. Recent encoder families such as GTE-MLM~\cite{gte_mlm}, ModernBERT~\cite{modernbert}, and Ettin-encoder~\cite{ettin} introduce stronger pretraining recipes and architectural improvements, including rotary positional embeddings and pre-normalization. Under the common drop-in-backbone assumption, these advances should translate into stronger SPLADE models~\cite{graniter2}.

\begin{figure}[t!]
\centering
\includegraphics[width=1.0\linewidth]{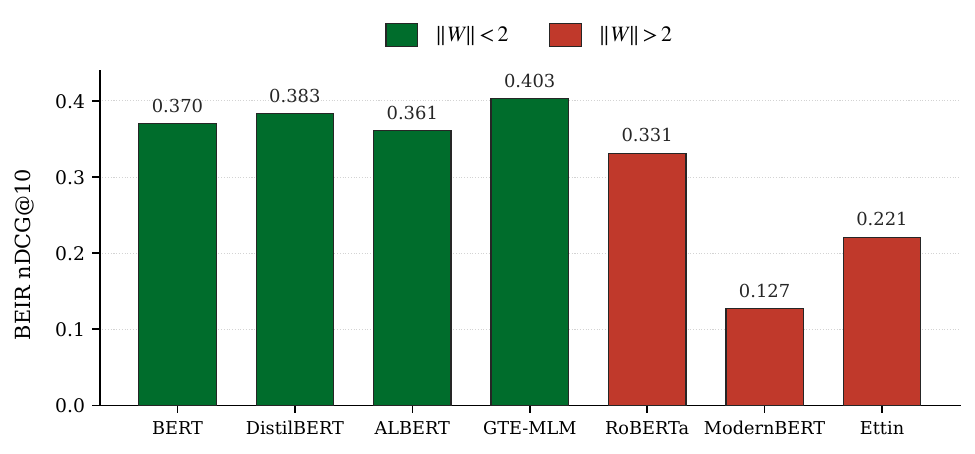}
\caption{MLM Head L2 norm $||W||$ and BEIR-13 effectiveness of MS MARCO-trained SPLADE models for each backbone.}
\label{fig:motivation}
\end{figure}

However, this assumption does not hold in practice. We train SPLADE models with different encoder backbones on MS MARCO~\cite{msmarco} using the same training recipe and evaluate them on BEIR-13. As shown in Figure~\ref{fig:motivation}, backbones with small MLM-head L2 norms ($||W||<2$), such as BERT, DistilBERT, ALBERT, and GTE-MLM, produce strong sparse retrievers. In contrast, backbones with larger norms ($||W||>2$), including RoBERTa, ModernBERT, and Ettin, substantially underperform the BERT baseline. This result is counterintuitive: stronger and more recent encoders can become weaker SPLADE backbones under the standard training recipe.

We identify the scale of the MLM head as a key factor behind this failure. In SPLADE, both query and document representations are produced through the MLM head and compared using an unnormalized sparse dot product. Therefore, an overly large MLM-head scale can amplify sparse activations, inflate matching scores, and distort the optimization dynamics of contrastive retrieval training.

Based on this observation, we propose a simple initialization-time correction: rescaling the MLM-head matrix by a constant factor $k$ before SPLADE training. This operation preserves the direction and lexical information of the pretrained vocabulary projection, introduces no additional parameters, requires no architectural modification, and adds no training or inference cost.

Despite its simplicity, this rescaling substantially improves SPLADE models built on encoders with large MLM-head norms. With $k=16$, ModernBERT achieves a $215\%$ relative improvement on BEIR-13 mean nDCG@10, while Ettin improves by $77\%$, relative to their unscaled counterparts. Across in-domain and out-of-domain benchmarks, we show that rescaled encoders recover strong SPLADE performance under the standard training recipe. These results suggest that MLM-head scale, rather than encoder capacity alone, is a critical factor for adapting pretrained encoders to learned sparse retrieval.

\section{RELATED WORK}
\label{sec:related_work}

Learned sparse retrieval represents queries and documents as sparse vectors over a vocabulary, combining the efficiency and interpretability of lexical retrieval with neural term weighting and expansion~\cite{splade, lsr}. This design preserves compatibility with inverted-index search, in contrast to dense retrievers that rely on low-dimensional continuous representations~\cite{e5, bgem3, qwen3embedding}, and has motivated recent work on efficient sparse indexing and retrieval systems~\cite{seismic, seismicv2, seismicv3}.

Among learned sparse retrieval methods, SPLADE has become a representative architecture. Subsequent variants have improved retrieval effectiveness through hard negative mining, knowledge distillation, sparsity regularization, and larger or more diverse training data~\cite{spladev3, milco, spladecode}. Other studies have analyzed the role of vocabularies and representation capacity in sparse retrieval models~\cite{rolevocab, exploringvocab, esplade}. However, these works largely preserve the core SPLADE formulation, where sparse lexical representations are produced through the pretrained vocabulary projection.

Backbone choice is another natural factor in SPLADE performance. SPLADE has traditionally relied on masked language models~\cite{bert}, while recent encoder families such as GTE-MLM~\cite{gte_mlm}, ModernBERT~\cite{modernbert}, and Ettin-encoder~\cite{ettin} provide stronger pretrained representations and appear to be natural replacements for BERT-style backbones. However, adapting these encoders to SPLADE also changes the vocabulary projection used for sparse retrieval.

\begin{table}[t!]
\centering
\caption{Backbone statistics. $||W||*{\mathrm{mean}}$ and $||W||*{\max}$
denote the mean and maximum row-wise L2 norm of the MLM-head.
}
\label{tab:model_stats}
\resizebox{\linewidth}{!}{%
\begin{tabular}{lcccc}
\toprule
\textbf{Model} & \textbf{\# Params} & \textbf{$||W||*{\mathrm{mean}}$} & \textbf{$||W||*{\max}$} & \textbf{Anisotropy} \\
\midrule
BERT-base          & 0.10B  & 1.401 & 2.045 & 0.445 \\
DistilBERT-base    & 0.07B  & 1.660 & 2.611 & 0.532 \\
ALBERT-base        & 0.011B & 0.407 & 0.737 & 0.108 \\
GTE-MLM-base       & 0.10B  & 1.278 & 1.814 & 0.085 \\
RoBERTa-base       & 0.10B  & 3.645 & 4.480 & 0.024 \\
ModernBERT-base    & 0.10B  & 2.553 & 4.273 & 0.044 \\
Ettin-encoder-150m & 0.15B  & 2.628 & 4.506 & 0.036 \\
\bottomrule
\end{tabular}
}
\end{table}

Table~\ref{tab:model_stats} summarizes the MLM-head statistics of the backbones considered in this work. The encoders differ not only in architecture and pretraining recipe, but also in the scale and geometry of their MLM-head matrices. While anisotropy has been discussed as an important property of representation geometry~\cite{anisotropy}, it does not align with the failure pattern observed in Figure~\ref{fig:motivation}: RoBERTa, ModernBERT and Ettin have relatively low anisotropy, yet perform poorly as SPLADE backbones under the standard training recipe. In contrast, the MLM-head norm provides a more consistent signal for the observed degradation. This motivates our focus on MLM-head scale as a previously overlooked factor in adapting modern encoders to learned sparse retrieval.

\section{METHOD}
\label{sec:method}

\subsection{SPLADE Training Objective}
\label{sec:splade_objective}

We briefly formalize SPLADE to isolate how the scale of the MLM head affects sparse activations, retrieval scores, and training dynamics. Let $W_{head} \in \mathbb{R}^{|\mathcal{V}| \times d}$ denote the MLM-head matrix that projects hidden states into vocabulary logits. Given contextualized token representations $H$, SPLADE constructs token-level vocabulary activations as
\begin{equation}
A = \log(1 + \mathrm{ReLU}(HW_{head}^{\top}))
\end{equation}
where the bias term is omitted for simplicity. The sequence-level sparse vector is then obtained by element-wise max pooling over token positions:
\begin{equation}
\mathbf{v} = \max_i A_i
\end{equation}
For a query $q$ and a document $d$, the retrieval score is computed by an unnormalized sparse dot product:
\begin{equation}
\mathrm{s}(q,d) = \mathbf{v}_q^{\top}\mathbf{v}_d
\end{equation}

Given query-positive-hard negative triples
$\{(q_i,d_i^+,d_i^-)\}_{i=1}^{B}$, we optimize an in-batch contrastive loss:
\begin{equation}
\mathcal{L}_{\mathrm{rank}}
=
-\frac{1}{B}
\sum_{i=1}^{B}
\log
\frac{
e^{(\mathrm{s}(q_i,d_i^+) / \tau)}
}{
\sum_{j=1}^{B}
\left[
e^{(\mathrm{s}(q_i,d_j^+) / \tau)}
+
e^{(\mathrm{s}(q_i,d_j^-) / \tau)}
\right]
}
\end{equation}
where $\tau$ is a temperature parameter, and $s$ denotes the similarity function (dot product). Following SPLADE, we additionally use FLOPS regularization~\cite{flops} to encourage sparse activations:
\begin{equation}
L = L_{\mathrm{rank}} + \lambda_q L_{\mathrm{FLOPS}}(\mathbf{v}_q) + \lambda_d L_{\mathrm{FLOPS}}(\mathbf{v}_d)
\end{equation}

\subsection{MLM-Head Rescaling}
\label{sec:mlm_head_rescaling}

The SPLADE objective makes the scale of the MLM head important. Since both query and document sparse vectors are produced from $HW^{\top}$ and then multiplied in an unnormalized dot product, the norm of $W$ can strongly affect the magnitude of sparse activations, retrieval scores, and gradients from the ranking loss. When the MLM-head scale is large, the ranking objective and FLOPS regularizer can be placed in sharper tension, making SPLADE training unstable or ineffective.
To control this factor, we apply a simple initialization-time correction before training:
\begin{equation}
W_{head} \leftarrow \frac{W_{head}}{k}
\end{equation}
Here, $k=1$ corresponds to the standard SPLADE initialization. For tied MLM heads, the operation is applied to the shared weight matrix so that weight tying is preserved. All other model parameters, architecture, training data, and objectives remain unchanged. Thus, the proposed correction introduces no additional parameters and requires no architectural modification.

\section{EXPERIMENTAL SETUP}
\label{sec:setup}

\subsection{Training Details}
\label{sec:training_details}

\paragraph{Backbone models}
We evaluate seven encoder backbones that cover both BERT-style and modern encoder families: bert-base-uncased, distilbert-base-uncased, albert-base-v2, gte-en-mlm, roberta-base, ModernBERT-base, and ettin-encoder-150m. These models differ in pretraining recipes, training data, architectural choices, and MLM-head statistics, allowing us to analyze whether SPLADE performance is determined primarily by encoder capacity or by the scale of the MLM-head projection. Unless otherwise specified, each backbone is trained under the same SPLADE recipe so that performance differences can be attributed to the backbone initialization and the proposed rescaling.

\paragraph{Data}
All SPLADE models are trained on MS MARCO triplets~\footnote{\url{https://huggingface.co/datasets/sentence-transformers/msmarco-msmarco-MiniLM-L6-v3}} using the same training objective, which combines a ranking loss with FLOPS regularization to encourage sparse representations. We keep the training recipe fixed across backbones, including the data, regularization setting, batch size, learning rate, and number of training steps. This controlled setup allows us to isolate the effect of MLM-head scale from other training factors.

\paragraph{Hyperparameters}
We optimize with AdamW at a peak learning rate of $1e-4$ under a linear schedule with a warmup ratio of $0.1$ for one epoch over the MS MARCO triplets. We use a per-GPU batch size of $64$ on $8$ NVIDIA RTX A6000 GPUs with Distributed Data Parallelism (DDP), a maximum sequence length of $256$, and bfloat16 mixed precision. The ranking loss uses in-batch negatives with temperature $\tau = 1.0$. For sparsity, we set the FLOPS regularizer weights to $\lambda_q = 5e-2$ and $\lambda_d = 3e-3$, ramped from zero with a quadratic schedule over the first third of training.

\subsection{Evaluation}
\label{sec:evaluation}

We evaluate retrieval effectiveness on both in-domain and out-of-domain benchmarks. For in-domain evaluation, we report results on MS MARCO dev~\cite{msmarco} and TREC-DL~2019~\cite{trecdl2019}. For out-of-domain evaluation, we use BEIR-13~\cite{beir} and report the mean nDCG@10 across tasks. We compare the standard initialization ($k=1$) with rescaled variants and analyze how the choice of $k$ affects SPLADE training and retrieval effectiveness.

\section{EXPERIMENTAL RESULTS AND ANALYSIS}
\label{sec:results}

\subsection{Main Result}
\label{sec:main_result}

\begin{table}[t]
\centering
\caption{Retrieval performance of SPLADE models with different MLM-head rescaling factors. The value in parentheses denotes the resulting MLM-head L2 norm $\|W\|$.}
\label{tab:main}
\resizebox{1.0\linewidth}{!}{%
\begin{tabular}{llcccc}
\toprule
\multirow{2}{*}{\textbf{Model}} & \multirow{2}{*}{\textbf{$k$ ($\|W\|$)}} & \multicolumn{2}{c}{\textbf{MSMARCO DEV}} & \textbf{TREC 2019} & \textbf{BEIR-13} \\
\cmidrule(lr){3-4} \cmidrule(lr){5-5} \cmidrule(lr){6-6}
& & \textbf{MRR@10} & \textbf{Recall@1000} & \textbf{nDCG@10} & \textbf{nDCG@10} \\
\midrule
BERT       & 1 (1.40) & .254 & .931 & .594 & .370 \\
DistilBERT & 1 (1.66) & .259 & .936 & .594 & .383 \\
ALBERT     & 1 (0.41)  & .255 & .926 & .530 & .361 \\
GTE-MLM    & 1 (1.28) & .273 & .953 & .635 & .403 \\
\midrule
\multirow{5}{*}{RoBERTa} 
           & 1 (3.65)  & .194 & .900 & .554 & .331 \\
           & 2 (1.82)  & .226 & \textbf{.934} & .590 & \textbf{.386} \\
           & 4 (0.91)   & \textbf{.228} & .933 & \textbf{.605} & .383 \\
           & 8 (0.46)   & .173 & .875 & .468 & .328 \\
           & 16 (0.23)  & .000 & .008 & .000 & .004 \\
\midrule
\multirow{5}{*}{ModernBERT} 
           & 1 (2.55)  & .045 & .554 & .266 & .127 \\
           & 2 (1.28)  & .201 & .909 & .502 & .351 \\
           & 4 (0.64)   & .214 & .925 & .570 & .373 \\
           & 8 (0.32)   & .248 & \textbf{.948} & .588 & \textbf{.405} \\
           & 16 (0.16)  & \textbf{.250} & \textbf{.948} & \textbf{.609} & .400 \\
\midrule
\multirow{5}{*}{Ettin} 
           & 1 (2.63)  & .099 & .752 & .354 & .221 \\
           & 2 (1.31)  & .118 & .777 & .416 & .244 \\
           & 4 (0.66)   & .175 & .890 & .530 & .332 \\
           & 8 (0.33)   & .178 & .899 & .538 & .330 \\
           & 16 (0.16)  & \textbf{.238} & \textbf{.940} & \textbf{.587} & \textbf{.391} \\
\bottomrule
\end{tabular}%
}
\end{table}

Table~\ref{tab:main} reports the retrieval performance of SPLADE models across MS MARCO Dev, TREC-DL~2019, and BEIR-13. We compare the standard SPLADE initialization ($k=1$) with MLM-head rescaling using different values of $k$. The number in parentheses indicates the resulting MLM-head L2 norm after rescaling.

First, performance under the standard SPLADE initialization is more closely aligned with MLM-head scale than with encoder capacity. Backbones with relatively small MLM-head norms, including BERT, DistilBERT, ALBERT, and GTE-MLM, consistently produce competitive sparse retrievers, with BEIR-13 mean nDCG@10 ranging from $0.361$ to $0.403$. In contrast, backbones with larger MLM-head norms substantially underperform: RoBERTa drops to $0.331$ on BEIR-13, while ModernBERT and Ettin collapse more severely to $0.127$ and $0.221$, respectively. This shows that stronger pretrained encoders are not necessarily reliable drop-in replacements for BERT in SPLADE when their MLM-head scale is not properly calibrated.
Second, MLM-head rescaling recovers the degraded large-norm backbones, but the effect is not monotonic. For ModernBERT, BEIR-13 nDCG@10 improves from $0.127$ to $0.405$, while MS MARCO MRR@10 improves from $0.045$ to $0.250$ and TREC-DL~2019 nDCG@10 from $0.266$ to $0.609$. Ettin also improves substantially, with BEIR-13 nDCG@10 increasing from $0.221$ to $0.391$ at $k=16$. At the same time, the optimal rescaling factor depends on the backbone: RoBERTa benefits from moderate rescaling and peaks at $k=2$, whereas ModernBERT and Ettin require stronger rescaling. Excessive down-scaling, however, can collapse performance, as shown by RoBERTa at $k=16$. These results indicate that the MLM-head scale must be controlled.

\begin{figure}[t!]
\centering
\includegraphics[width=1.0\linewidth]{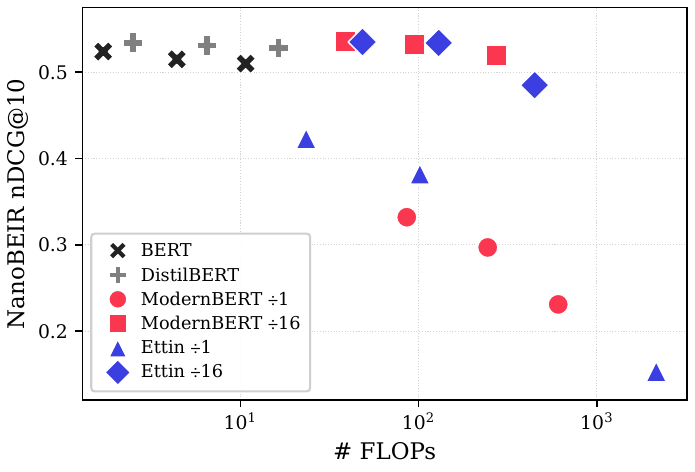}
\caption{Effectiveness--sparsity trade-off on NanoBEIR under different FLOPS
regularization strengths. 
}
\label{fig:flops_tradeoff}
\end{figure}

\subsection{Effectiveness--Sparsity Trade-off}
\label{sec:flops_analysis}

We next examine whether rescaling improves not only retrieval effectiveness but
also the sparsity--effectiveness trade-off. For this analysis, we vary the
FLOPS regularization weights over
$\lambda_q \in \{1e-2, 5e-2, 2e-1\}$ and
$\lambda_d \in \{6e-4, 3e-3, 1e-2\}$, while keeping
the other training settings fixed. Figure~\ref{fig:flops_tradeoff} plots
NanoBEIR mean nDCG@10 against the number of activated FLOPs. Successful SPLADE
models should achieve high effectiveness while keeping FLOPs low, since sparse
retrieval depends on compact inverted-index representations.

Without rescaling, ModernBERT and Ettin exhibit an unfavorable trade-off: they
often require substantially more FLOPs than BERT and DistilBERT while achieving
lower NanoBEIR effectiveness. After MLM-head rescaling, both models move toward
the high-effectiveness, low-FLOPs region occupied by successful SPLADE
backbones. This suggests that rescaling improves the effectiveness--sparsity trade-off by producing better-calibrated sparse activations, rather than simply increasing the density of the representations.

\subsection{Rescaling Requires an Appropriate Scale}
\label{sec:scaling_range}

\begin{figure}[t!]
\centering
\includegraphics[width=1.0\linewidth]{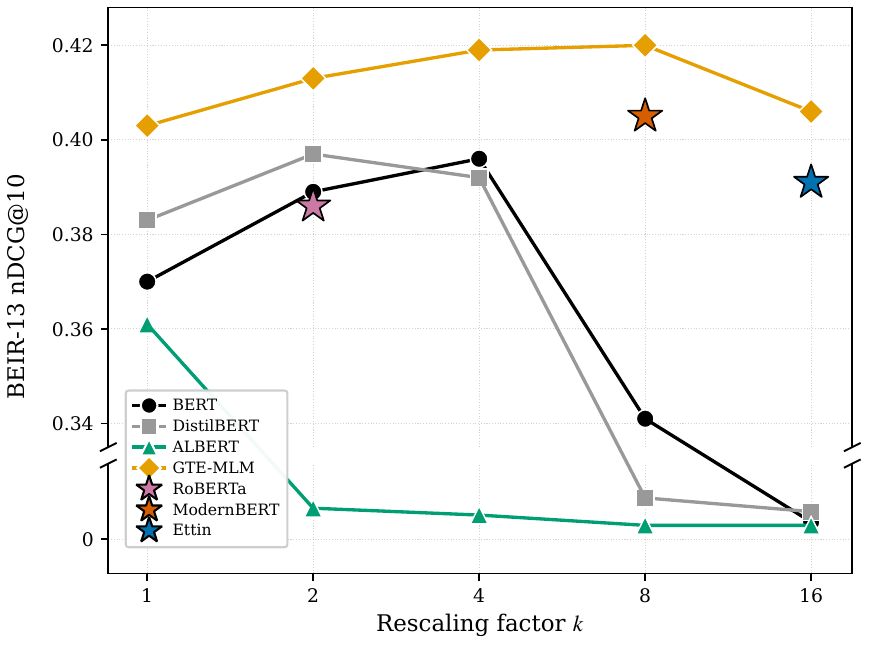}
\caption{Effect of MLM-head rescaling on BEIR. Lines show small-norm backbones across rescaling factors, while stars indicate the best rescaled results for large-norm backbones. 
}
\label{fig:scaling_range}
\end{figure}

We also test whether rescaling is simply a universally beneficial
regularization trick. Figure~\ref{fig:scaling_range} shows BEIR-13 nDCG@10 as
the rescaling factor varies for backbones whose initial MLM-head norms are
already relatively small. Moderate rescaling can slightly improve BERT,
DistilBERT, and GTE-MLM, but excessive rescaling consistently degrades
performance. ALBERT is the clearest case: because its initial MLM-head norm is
already very small, further down-scaling rapidly collapses retrieval
effectiveness.

These results show that the benefit of rescaling does not come from shrinking
the MLM head as much as possible. Instead, SPLADE training requires the
MLM-head scale to lie within an appropriate range. Large-norm backbones such as
ModernBERT and Ettin benefit from stronger rescaling because their initial
projection scale is too large, whereas already small-norm backbones can be
harmed when the projection is scaled down too aggressively. Together with the main results, this
suggests that MLM-head scale is a critical but previously overlooked factor in
adapting pretrained encoders to learned sparse retrieval.

\subsection{Rescaling Stabilizes Training}
\label{sec:loss_analysis}

\begin{figure}[b]
\centering
\includegraphics[width=1.0\linewidth]{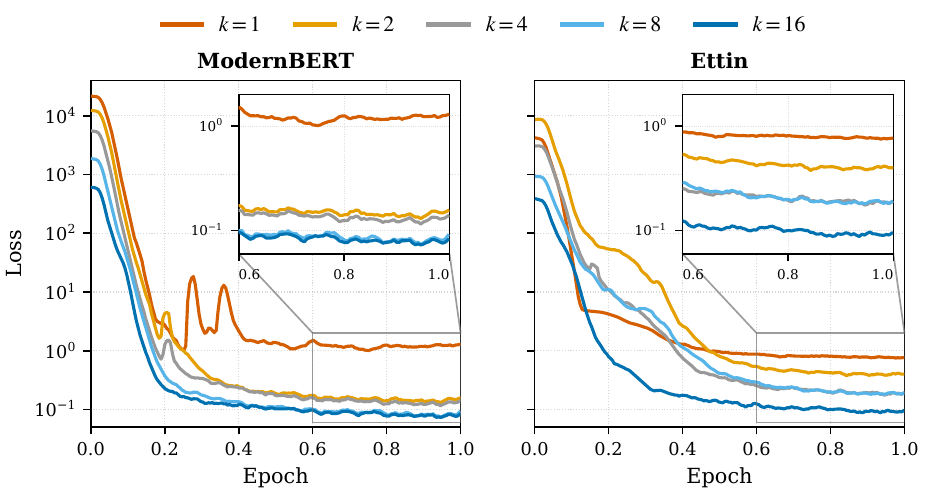}
\caption{Training loss curves of ModernBERT- and Ettin-based SPLADE models
under different rescaling factors.}
\label{fig:loss_analysis}
\end{figure}

Finally, Figure~\ref{fig:loss_analysis} analyzes the training dynamics of the two
large-norm modern encoders. Without rescaling ($k=1$), ModernBERT and Ettin
start from extremely large losses and converge to substantially higher final
losses than their rescaled variants. As $k$ increases, the loss scale is reduced
and training becomes smoother, indicating that MLM-head rescaling directly
stabilizes the optimization process. This supports our interpretation that
large MLM-head norms distort SPLADE training by inflating the score scale in the
contrastive objective, while rescaling moves modern encoders into a more
appropriate optimization regime.

\section{CONCLUSION}

In this work, we showed that encoder backbones with large MLM-head scales can fail under the standard SPLADE training recipe, even when the underlying pretrained encoders are strong. Through experiments across multiple backbones, we identified the MLM-head L2 norm as a key factor associated with this failure: large-norm MLM heads amplify sparse activations and destabilize retrieval training, whereas appropriately scaled heads lead to more effective sparse retrievers. To address this issue, we introduced a simple initialization-time rescaling correction that adjusts the MLM-head matrix before training. This zero-cost modification restores the effectiveness--sparsity trade-off of large-norm SPLADE backbones and substantially improves both in-domain and out-of-domain retrieval performance. Our findings suggest that adapting pretrained encoders to learned sparse retrieval requires not only stronger representations, but also careful control of the projection scale used to construct sparse lexical vectors.


\section*{GenAI Usage Disclosure}
ChatGPT (GPT-5.5) was used to check spelling and improve grammar during the writing of this paper.

\begin{acks}
To Robert, for the bagels and explaining CMYK and color spaces.
\end{acks}

\bibliographystyle{ACM-Reference-Format}
\bibliography{custom}

\appendix









\end{document}